\documentclass[prd,twocolumn,superscriptaddress]{revtex4}

\usepackage{graphics}

\begin{document}

\title{Cosmic String Induced CMB Maps }

\author{M. Landriau}
\email[Email: ]{landriau@mpe.mpg.de}
\affiliation{Max-Planck-Institut f\"{u}r extraterrestrische Physik,
Giessenbachstra{\ss}e 1, 85748 Garching, Germany}
\affiliation{Department of Applied Mathematics and Theoretical Physics,
University of Cambridge,
Centre for Mathematical Sciences, Wilberforce Road, Cambridge CB3 0WA,
United Kingdom}
\author{E.P.S. Shellard}
\email[Email: ]{epss@damtp.cam.ac.uk}
\affiliation{Department of Applied Mathematics and Theoretical Physics,
University of Cambridge,
Centre for Mathematical Sciences, Wilberforce Road, Cambridge CB3 0WA,
United Kingdom}

\date{\today}

\begin{abstract}

We compute maps of CMB temperature fluctuations seeded by cosmic
strings using high resolution simulations of cosmic strings in a
Friedmann-Robertson-Walker universe.   We create full-sky, 18$^\circ$ and $3^\circ$ CMB maps, 
including the relevant string contribution at each resolution from before recombination
to today.  We extract the angular power spectrum from these maps, demonstrating
the importance of recombination effects.  We briefly discuss the probability density 
function of the pixel temperatures, their skewness,  and kurtosis. 

\end{abstract}

\maketitle

\section{Introduction}

Despite improving observational limits, interest in cosmic strings 
has remained durable (for a review, see~\cite{vilenkin94}).  Strings are 
a generic phenomena in fundamental theories and they can emerge in 
macroscopic form in braneworld cosmologies, for example, at the end 
of inflation~\cite{sarangi2002,copeland2004}.   They are also common
to cosmologically viable supersymmetry grand unified theory models~\cite{jeannerot2003}.
Stringent constraints on strings are important, therefore, in restricting 
the latitude available for cosmological model building.  The detection
of cosmic strings would be a watershed for high energy theory.   

Despite the potential significance, the investigation of cosmic
strings and their observational consequences faces many numerical and
analytic challenges, not least in creating accurate realizations of
string imprints in the cosmic microwave sky.  In this paper, we take
this study a step further forward by presenting full-sky and
small-angle CMB maps of temperature fluctuations seeded by cosmic
string networks using high resolution simulations in an
Friedmann-Robertson-Walker expanding
universe (with the longest dynamic range to date).  This work includes
all the relevant recombination physics and can be used not only to
determine the angular power spectrum of string CMB anisotropies but
also the higher order correlators such as the bispectrum, trispectrum,
and beyond.

Current constraints on cosmic strings result from line-of-sight CMB
power spectrum calculations sourced either by unequal-time correlators
obtained from field theory string
simulations~\cite{pen97,durrer99,bevis2004} or semianalytic models of
Nambu strings~\cite{battye98,pogosian99}.  Qualitatively these two
approaches produce consistent spectra, that is, without the strong
coherent acoustic peaks associated with inflation.  However,
quantitatively there is a mismatch between the two approaches in both
the shape of the primary peak and its amplitude, which differs by a
factor of 2--3.  This disparity arises primarily from a difference in
string network densities, which has been discussed at some length
elsewhere~\cite{martins2004} (see also \cite{battye2008}).
Nevertheless, there is general agreement that the relative amplitude
of string induced CMB fluctuations cannot exceed more than 10\% of
those arising from adiabatic inflationary perturbations
\cite{battye98,bevis2004}.  There have also been a number of studies
going beyond the power spectrum through map making with cosmic
strings~\cite{bennett88, allen96, lacmb, fraisse2008} in order to
study the degree of Gaussianity of the resulting CMB signatures.
However, this work has generally only included late-time gravitational
effects, ignoring the recombination physics which makes an important
contribution to the signal over a wide range of multipoles $l\approx
200$-$2000$.

The motivation for the present work, then, is twofold: first, to
include all recombination effects in the string CMB maps, so that we
can ultimately characterize their primary statistical properties,
and secondly, to match the accuracy of future experiments such as
Planck~\cite{bluebook}, AMI~\cite{AMI2008} and QUIET~\cite{QUIET2007}
which will impose considerably more stringent constraints on cosmic
strings through improving precision, resolution, and added polarization
information.

\section{Cosmic string network simulations}

Cosmic string simulations were performed with the Allen-Shellard
string network code~\cite{allen90}.  We have used fixed comoving
resolution together with an initial string resolution of 24 points per
correlation length.  Simulations which started in the matter era had a
dynamic range of 7.5 in conformal time, i.e. $\eta_f = 7.5\eta_i$, but
in order to use only the simulation when the network has relaxed into
a scaling regime, we ignored the first 4\% of time steps, resulting
in an effective dynamic range of 6 in conformal time.  Simulations
that started in the radiation era had an effective dynamic range of 5,
after eliminating the first 4\% of time steps from a simulation with a
dynamic range of 6.  The energy-momentum tensor of the network was
projected onto a grid of $256^3$ points as described
in~\cite{methods}.  The background cosmology used was $\Lambda$CDM
with the WMAP 5-year data best fit parameters~\cite{wmap5params}:
$\Omega_{CDM} = $0.214, $\Omega_b = $0.044, $\Omega_{\Lambda}$ = 0.742,
and $h$ = 0.719.

We have used three string simulations which span the range from before equality to today.  
The epochs of each simulation are (1) from $\eta = \eta_0/7.5$ to
$\eta_0$, (2) from $\eta = \eta_0/45$ to $\eta_0/6$, and (3) from $\eta = \eta_0/216$ to $\eta_0/36$.  
Thus, these simulations together span a range of 180 in conformal time (note that since we
ignore the beginning of each simulation,  they do not overlap).

\section{Einstein-Boltzmann evolution}

We use the Landriau-Shellard code~\cite{methods} to compute
cosmological perturbations in Fourier space.  Two changes have been made to this code
since the algorithm was presented.

The first modification concerns the scalar metric equations employed: Instead of solving for $\dot{h}$ and $\dot{h}^S$,
we now solve for $\dot{h}$ and $\dot{h}^-\equiv \dot{h}-\dot{h}^S$, which obeys the following equation:
\begin{equation}
  \ddot{h}^- + 2\frac{\dot{a}}{a}\dot{h}^- = -16\pi G(p\Sigma + \delta p) + 8\pi G(\Theta + 2\Theta^S)
\end{equation}

The other modification concerns the inverse computation of the evolution equations' fundamental matrices:
Instead
of LU factorization, we use singular value decomposition, for which we employ the freely available LAPACK routines.
This has proved a more numerically stable method and enables a better treatment of near singular matrices~\cite{NRiC2},
especially around recombination.

\section{Maps}

We compute maps of CMB fluctuations by following photon paths through the simulation boxes.
The CMB temperature fluctuations are given by the following equation, obtained by integrating
the linear Boltzmann equation for the Stokes parameter $I$:
\begin{equation}\label{eq:tcmb}
\begin{array}{ll}
\displaystyle\frac{\delta T}{T} = \int_0^{\eta_0} &
\left(\dot{\tau}e^{-\tau} \left(\frac{1}{4}\delta_{\gamma} -
\mathbf{v_B}\cdot\hat{n} + \frac14\Pi^I_{ij}\hat{n}_i\hat{n}_j\right)
\right.\\
& \left. -\frac{1}{2}e^{-\tau}\dot{h}_{ij}\hat{n}_i\hat{n}_j
\right)d\eta
\end{array}
\end{equation}
where $\Pi^I_{ij}$ is the term that couples the Stokes parameter $I$ to $Q$ and $U$ and is
given in Fourier space by
\begin{equation}
\begin{array}{ll}\displaystyle
\Pi^I_{ij}(\mathbf{k}) & = \frac{3}{4}\left(\hat{k}_i\hat{k}_j-\frac13\delta_{ij}\right)\Pi^S\\
&
-\frac{1}{2}\left((\hat{k}_i\hat{e}_{1j}+\hat{k}_j\hat{e}_{1i})\Pi^{V1}
+(\hat{k}_i\hat{e}_{2j}+\hat{k}_j\hat{e}_{2i})\Pi^{V2}\right)\\
&
+(\hat{e}_{1i}\hat{e}_{1j} - \hat{e}_{2i}\hat{e}_{2j})\Pi^{T+} 
+(\hat{e}_{1i}\hat{e}_{2j} + \hat{e}_{2i}\hat{e}_{1j})\Pi^{T\times}
\end{array}
\end{equation}
and all the other terms have their usual meaning.  A full derivation of this formula
is given in~\cite{pol}.

In practice, because $\dot{\tau}e^{-\tau}$ and $e^{-\tau}$ are zero before the start of
recombination, we only output grids for all perturbations from $\eta = 2\eta_{rec}/3$ to the end
of simulation 3, which is $\eta \simeq 3\eta_{rec}/2$.  For simulations 1 and 2, we only output
the grids for $\dot{h}_{ij}$, because $\dot{\tau}e^{-\tau}$ is also zero after the end of recombination.

\begin{figure*}
\centering
\resizebox{!}{5cm}{\rotatebox{90}{\includegraphics{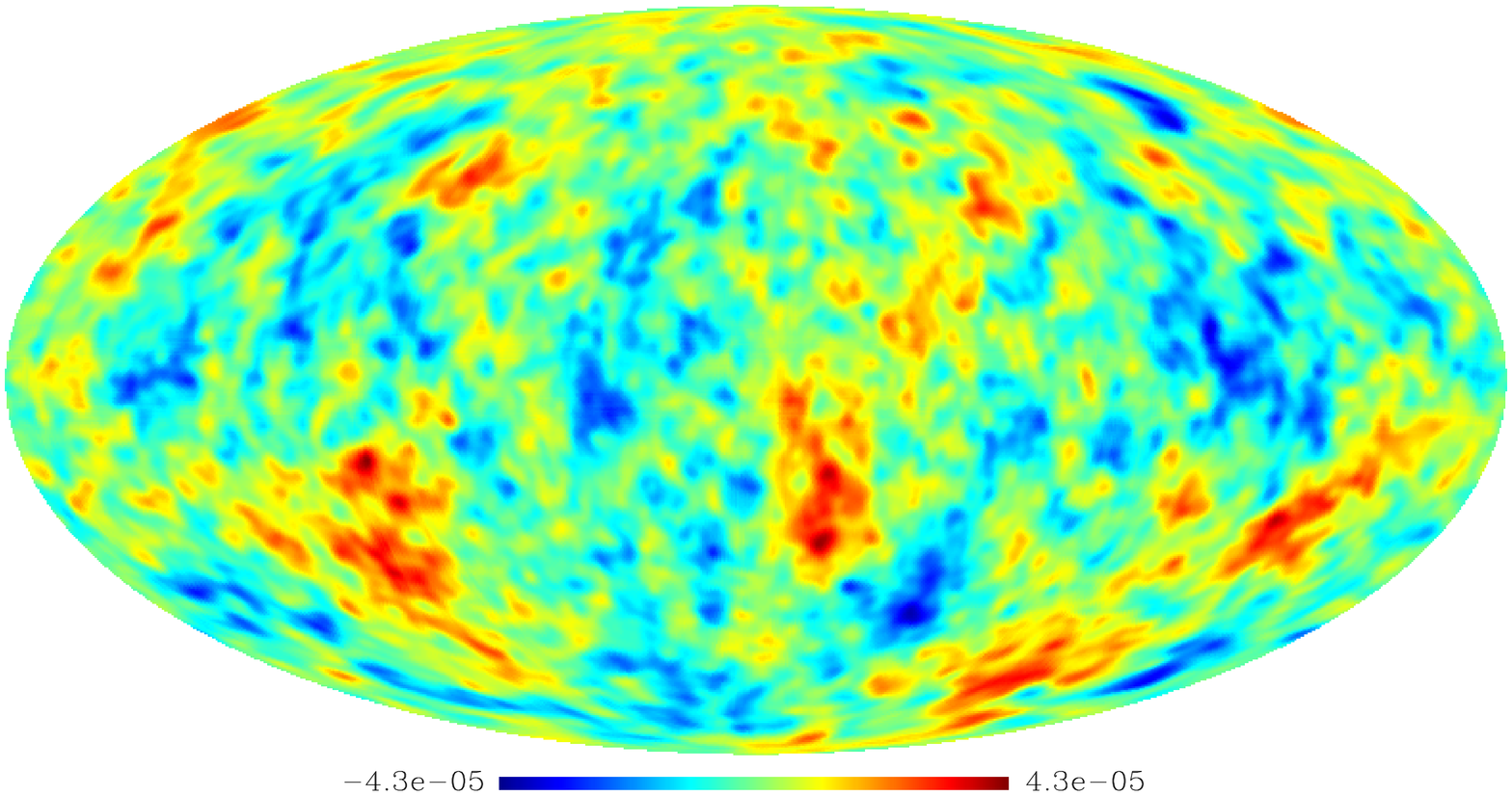}}}
\resizebox{4cm}{!}{\rotatebox{90}{\includegraphics{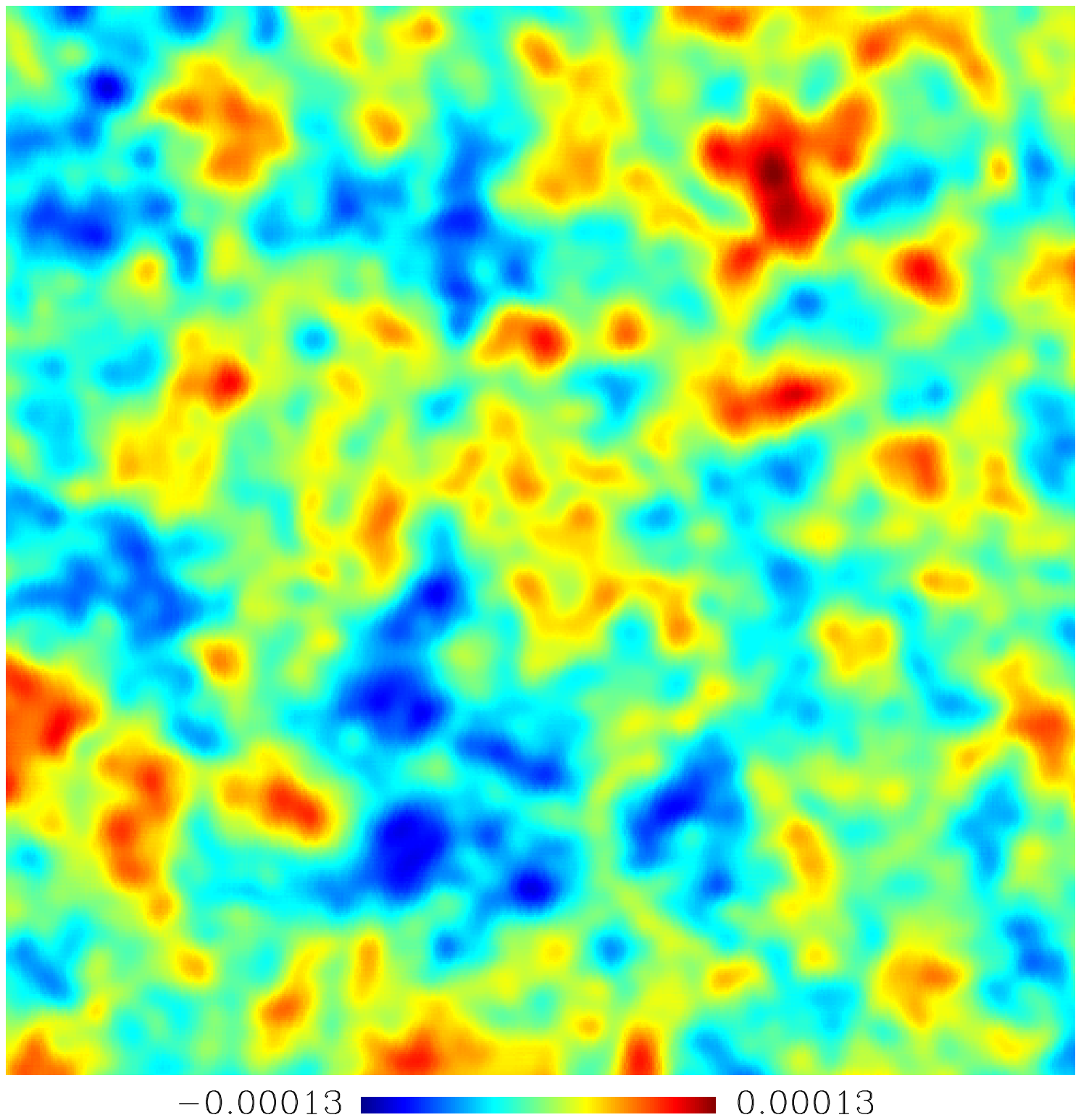}}}
\resizebox{4cm}{!}{\rotatebox{90}{\includegraphics{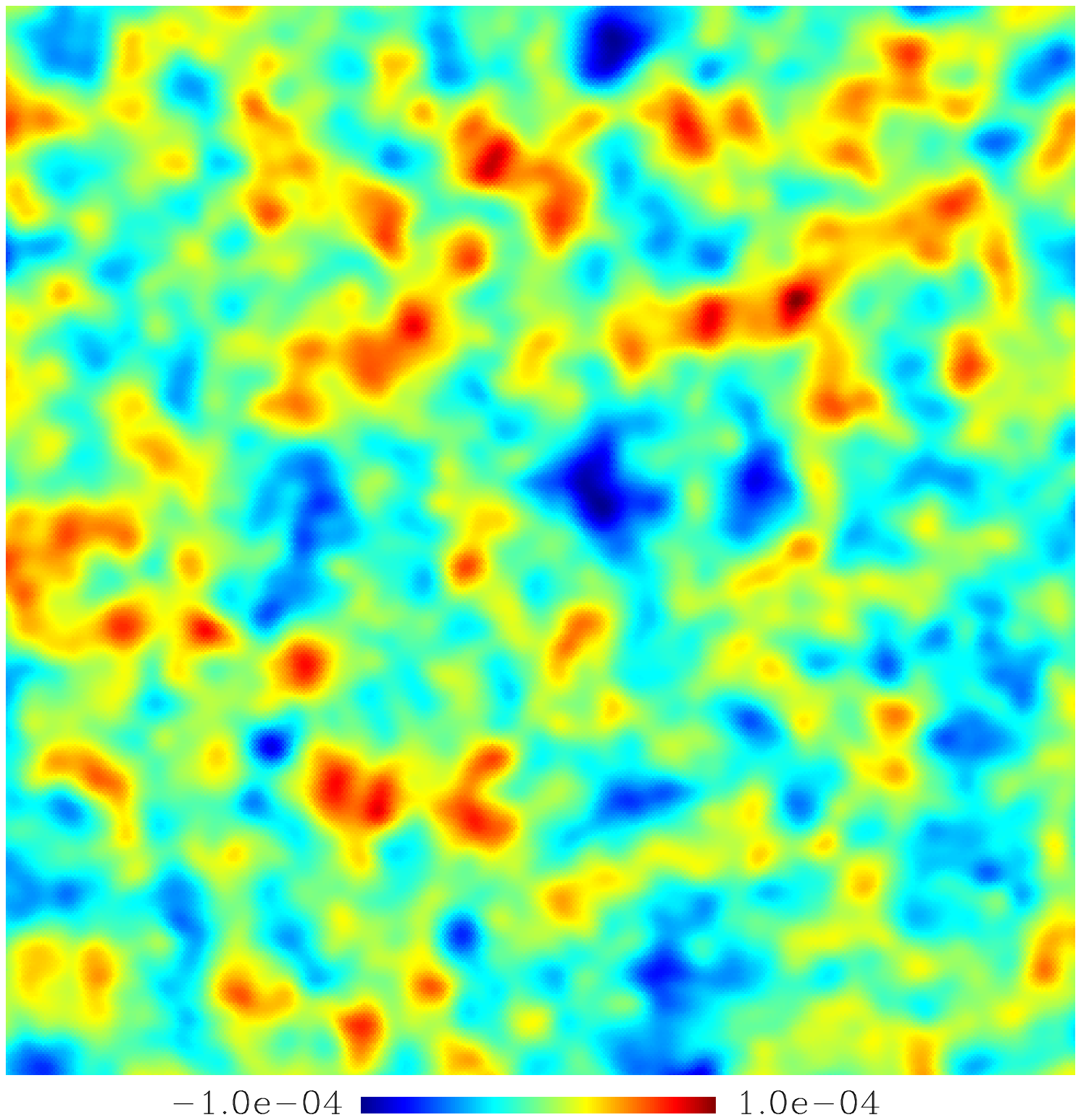}}}
\caption{Temperature fluctuations produced by cosmic strings from
 simulation 1 (left), simulation 2 (middle),
and simulation 3 (right).}\label{maps}
\end{figure*}

By putting ``observers'' at each apex of a cube of side $L/2$, where
$L$ is the simulation box size, we produce eight all-sky maps of resolution $N_{side}=256$
from simulation 1.
From simulation 2, we compute six $18^{\circ}\times 18^{\circ}$ maps of
resolution of $N_{side}=2048$, by putting an observer outside each
face of the simulation box.
Finally, from simulation 3, using the same setup as for simulation 2,
we compute six $3^{\circ}\times 3^{\circ}$ maps of $N_{side}=8192$.
Figure~\ref{maps} shows one map produced from each simulation;
it should be noted that even the maps of patches of sky are
computed using a spherical sky.

Comparing our maps with those of~\cite{fraisse2008}, we note that the
$18^{\circ}$ map shown and their $7.2^{\circ}$ map have similar features,
but the former does not present as sharp linelike discontinuities as
the latter.  The resolution
of our maps is effectively lower than that implied by the Healpix
$N_{side}$ parameter used.  This can be seen most directly from the
power spectra (see Sec.~\ref{sec:cls}): Normally, one would expect
$\ell_{max} \lesssim 3N_{side}$, but the power in the maps falls
around $\ell \approx N_{side}$, which shows they are over pixelized or,
to put it another way, the effective resolution is about a third of that
expected from the pixelization.  For example, our $18^{\circ}$ maps
have an effective resolution of about $5^{\prime}$, compared to
$1.7^{\prime}$ expected from a map of $N_{side}=2048$.  This explains
the difference in features with the Fraisse et al. maps, which have a
resolution of $0.42^{\prime}$, an order of magnitude higher than ours.

\section{Power spectrum computation}\label{sec:cls}

\begin{figure*}
\resizebox{\columnwidth}{5cm}{\includegraphics{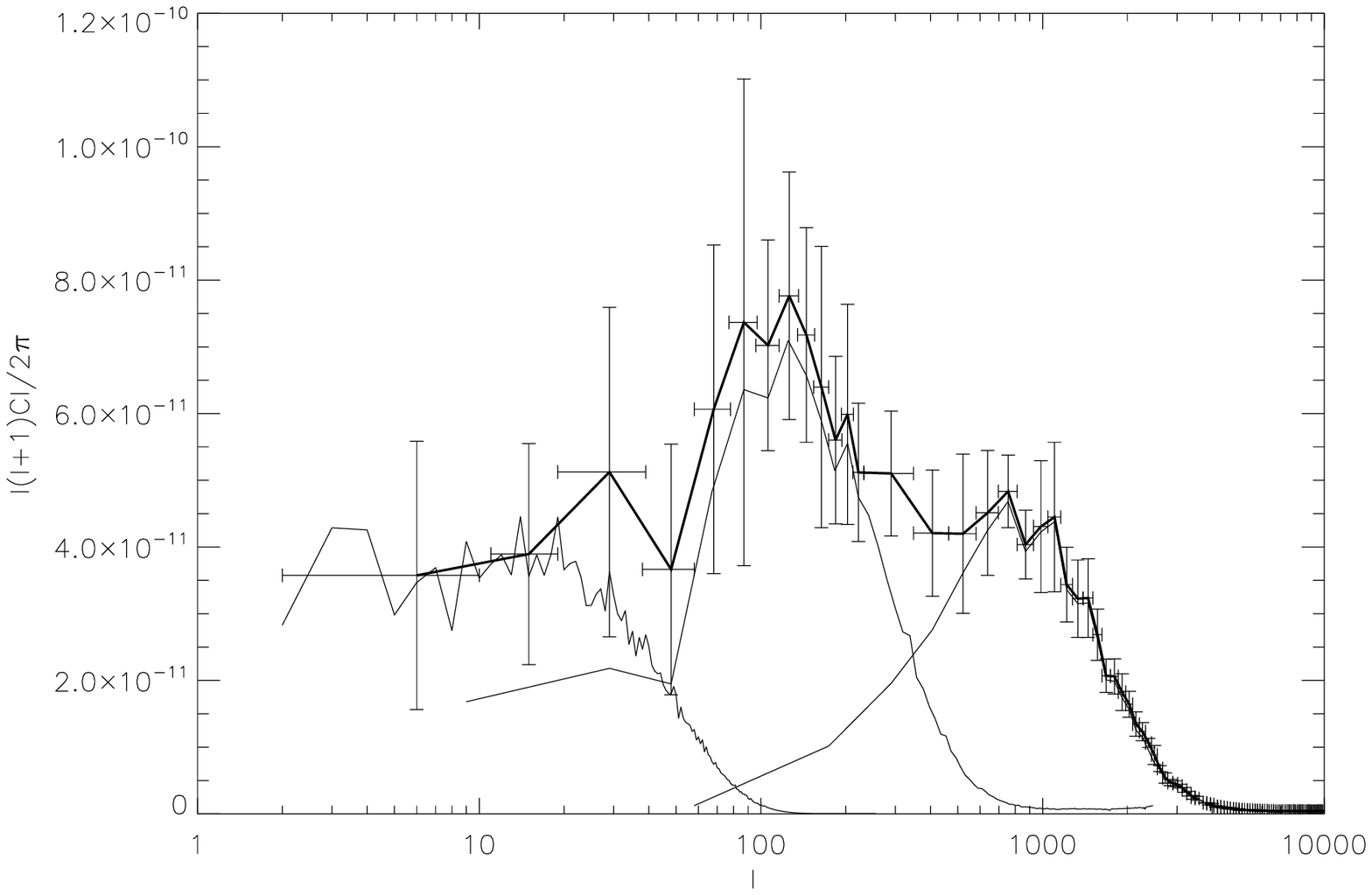}}
\resizebox{\columnwidth}{5cm}{\includegraphics{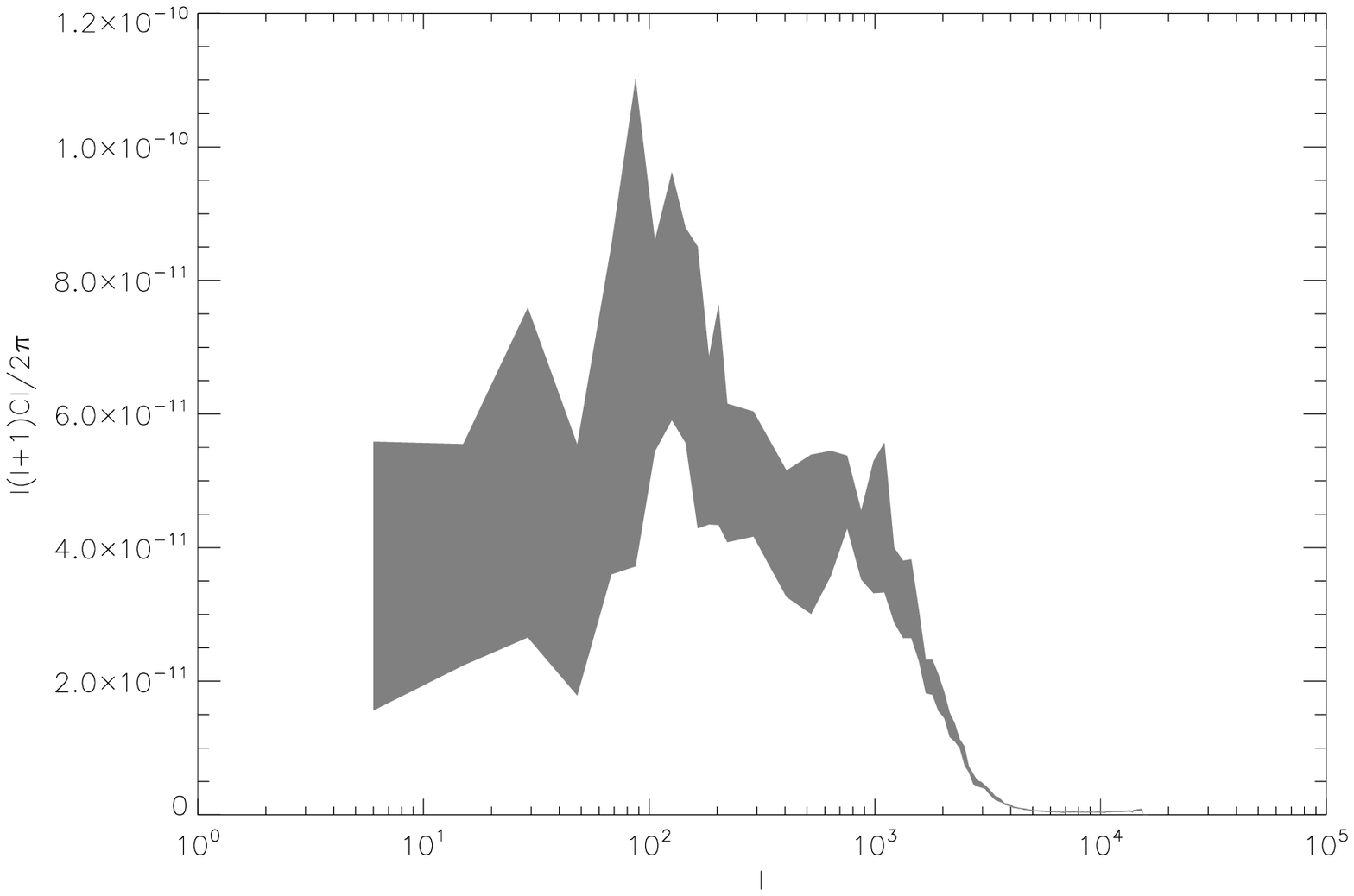}}
\caption{On the left, we show the angular power spectrum of CMB temperature fluctuations
  produced by networks of cosmic strings  The thin lines represent
  the spectra from simulations 1 -- 3 (left to right) and the
  thick line is their sum.  On the right, the total spectrum with its error
  bars is shown again as an ``expectation area.''
  }\label{fig:cl}
\end{figure*}

For the all-sky maps, we decompose the temperature fluctuations in the sky in
spherical harmonics: $\frac{\delta T}{T} = \sum a_{\ell m}Y_{\ell}^m$ 
and the angular power spectrum is estimated from
\begin{equation}
C_{\ell} = \frac{1}{2\ell +1}\sum_m a_{\ell m}a^{*}_{\ell m}\,.
\end{equation}
For this purpose, we use the Healpix package~\cite{gorski2005}.
For the $3^{\circ}$ patches, 
we use the flat-sky approximation (see e.g.~\cite{white99}) which
replaces the spherical harmonic transform with a 2-D Fourier
transform: $C_{\ell} \simeq C_{k} = |a_{\mathbf{k}}|^2$,
where the modes are obtained from $\frac{\delta T}{T} = \sum
a_{\mathbf{k}} e^{i\mathbf{k}.\mathbf{x}}$.
For the $18^{\circ}$ patches,
we have used both methods to compare their relative merit.  To use a
spherical transform on an incomplete sky, one must multiply the
extracted spectrum with a mode decoupling matrix~\cite{hivon2002}:
$C_{\ell} = M_{\ell\ell^{\prime}}^{-1}\tilde C_{\ell^{\prime}}$,
where
\begin{equation}
M_{\ell\ell^{\prime}} = (2\ell^{\prime}+1)\sum_L\frac{(2L+1)}{4\pi}C_{L}^{mask}
\left(\begin{array}{ccc} \ell & \ell^{\prime} & L \\ 0 & 0 & 0 \end{array}\right
)^2\,,
\end{equation}
However, the resulting spectra show strong oscillations due to the Gibbs
effect that are not corrected by the procedure outlined above.  Because
of this, we have used a 2D Fourier transform even though the flat-sky
approximation starts to break down at the lower end of the multipole
range probed by these maps.
In Fig.~\ref{fig:cl}, we show the angular power spectrum from each
of the three simulation sets, as well as their summation
(i.e. concatenating these consecutive but nonoverlapping time
contributions).  For the full sky, all multipoles are shown, while for
the patches of sky, the spectra are binned.  The errors on the
individual parts of the spectrum are estimated by the variance between
the maps.  To add up the spectra in sections where two sets of maps
contribute, we average the lower $l$ part in bins of the same size as
that of the higher $l$ part and then add the two contributions.  Also,
to reduce the error, we binned the part of the spectrum in which only
the all-sky maps contribute.  We normalize the string spectra using
$G\mu/c^2 = 1\times 10^{-6}$.

\begin{figure}
\resizebox{\columnwidth}{5cm}{\includegraphics{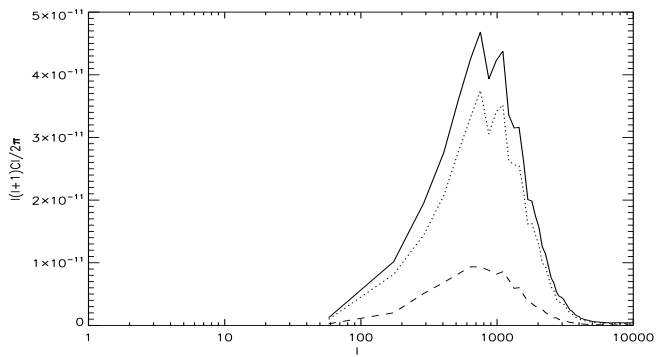}}
\caption{Power spectrum from simulation 3 using the full simulation
  (solid line) and the late start evolution (dashed line) and their
  difference (dotted line).}\label{fig:rec_effects}
\end{figure}

In addition, 
Fig.~\ref{fig:rec_effects} illustrates the power spectrum from simulation 3
which begins before equal matter radiation and separates out the
late-time contribution (dashed line), that is, as if the string
simulation and Einstein-Boltzmann evolution were to start just before
decoupling $\eta_{\rm dec}$.  Note that maps by other groups have been
generated by considering contributions only after decoupling
$\eta>\eta_{\rm dec}$, e.g.~\cite{fraisse2008}.  Here, we see starkly
illustrated the importance of the early contributions to the CMB
anisotropy which arise from matter and radiation perturbations induced
by strings before decoupling (primarily scalar modes).  For $l>400$,
this strong early scalar component predominates over the late-time
gravitational string contribution (primarily vector modes).  These
results contrast markedly with the subdominant scalar power spectra
obtained earlier using Nambu string simulations with full
Einstein-Boltzmann evolution in a flat cold dark matter model~\cite{allen97}.
However, this enhancement is due here to inclusion of the full
matter-radiation transition with a significantly higher radiation era
string density and a $\Lambda$ concordance cosmology with a higher
relative baryon density at decoupling.
In future work~\cite{uetc}, we shall compare this result with the
spectrum obtained from the same simulations' UETCs.
The overall shape for the spectrum computed from the $18^{\circ}$ maps
is also qualitatively consistent with the CMB power spectrum calculated
using a simplified analytic formula~\cite{RS09}.

Comparing this result with that of~\cite{fraisse2008} -- the only other
group to have recently extracted $C_{\ell}$ from simulated maps -- 
we see some significant differences attributable both
to the inclusion of recombination physics in
our work and to the different approximations used. Firstly,
we note that Ref.~\cite{fraisse2008} achieves much better statistics
because it is easier to simulate purely gravitational effects
in a flat-sky approximation at high resolution, generating a
larger number of maps which results in smaller error bars.
However, for $l\le 500$,
we note that their power spectrum does not appear to
turn over in the range $l\lesssim 200$--$ 400$, a causality cut-off
due to the fact that these string modes are superhorizon
at recombination (see e.g. Ref.~\cite{RS09}).  In the range $l\approx 400$--$2000$,
as discussed above, there is an important contribution from
early-time scalar modes at decoupling which is not included
in Ref.~\cite{fraisse2008}.   It is reasonable to assume that late-time
gravitational effects dominate the spectrum at $l \gg 1000$,
and their result in this range is probably fairly accurate.
However, in this region we note that the limited spatial resolution of
our individual maps (resulting from propagation of photons
through strings smoothed onto $256^3$
grids) means that there is generically small-scale power
missing at higher multipoles from the summed power
spectrum in Fig.~\ref{fig:cl}. For example, the gravitational contribution from strings in simulation 1 (full-sky) is absent
beyond $l\approx 100$. For this reason
 and given the limited number of realizations, 
we have not yet endeavored to obtain a detailed constraint using WMAP,
though we expect it to be consistent with constraints in Ref.~\cite{battye2010}.
By comparing the Sachs-Wolfe plateau to CMB observations, we find a
normalization of the cosmic strings linear energy density to be $G\mu
/c^2 = (1.45 \pm 0.6)\times 10^{-6}$, in agreement with our previous
result~\cite{lacmb}.  

\section{Probability distribution function}

One of the principal aspects of cosmic string induced CMB fluctuations
is their intrinsic non-Gaussianities.  In this section, we present preliminary 
results from our map realizations for the 
pixel temperature distribution, its skewness and kurtosis.
In future work, we shall explore the efficiency of different
techniques and tools, such as the bispectrum~\cite{bispec}, to better characterize the non-Gaussian
signature from strings and thus infer their detectability.

\begin{figure*}
\resizebox{\columnwidth}{5cm}{\includegraphics{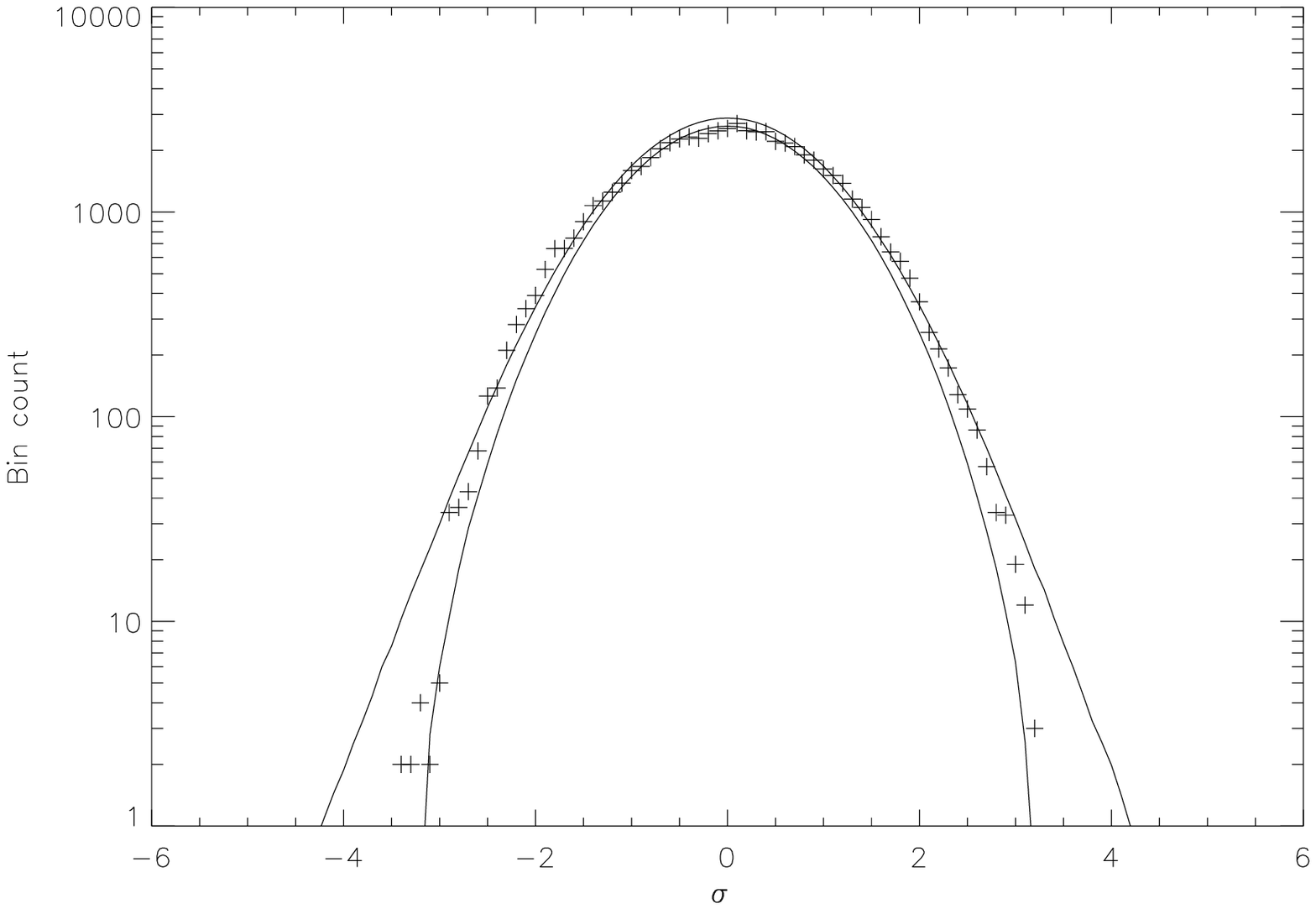}}
\resizebox{\columnwidth}{5cm}{\includegraphics{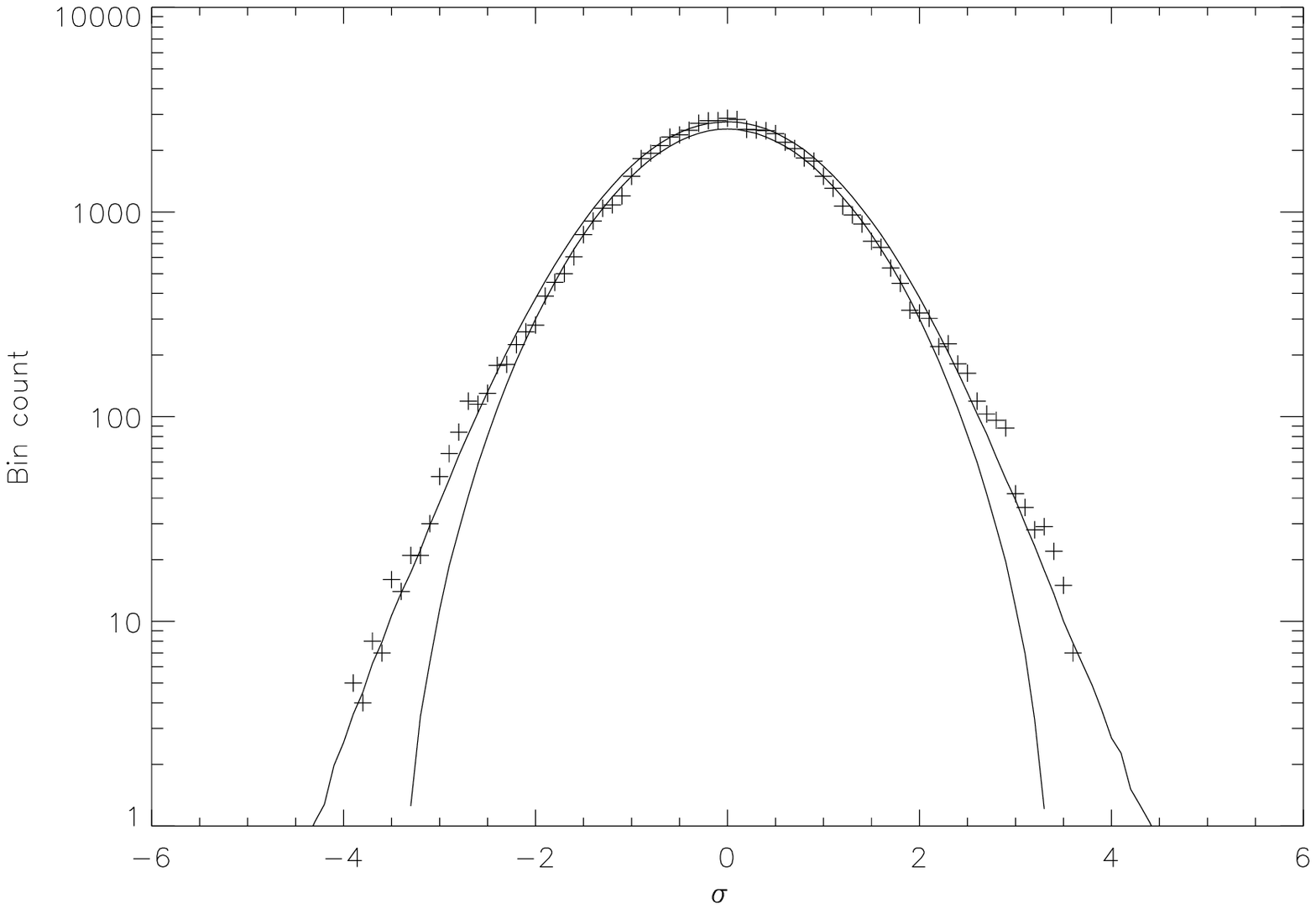}}
\caption{Temperature distributions of the medium- (left) and small- (right) angle maps shown above.
The solid lines indicate the $1\sigma$ level of
  an ensemble of Gaussian maps}\label{fig:ng}
\end{figure*}

We compute the pixel probability distribution function of the rectangular map pixel grids computed
to estimate the power spectrum.
To compare our results, we generate an ensemble of 1000 Gaussian maps with the same power
spectrum as the string map.  To do so, we use the fact that, in Fourier space, a Gaussian field
will have phases that are uniformly distributed between $0$ and $2\pi$.  Hence, for a given map,
after fast Fourier transform, we randomize the phases and fast Fourier
transform back to real space.
In Fig.~\ref{fig:ng}, we show
the temperature distribution for medium- and small-angle maps along with the Gaussian ensemble for
comparison.
These probability distribution functions are remarkably Gaussian, consistent with early string map results which demonstrated that, 
despite the distinct signature of individual strings, the central limit theorem prevailed after many strings
contributed~\cite{allen96}.  
We note, however,  that the higher resolution $3^\circ$ maps do appear to have 
a slightly increased level of non-Gaussianity. 

\begin{figure*}
\resizebox{\textwidth}{5cm}{\includegraphics{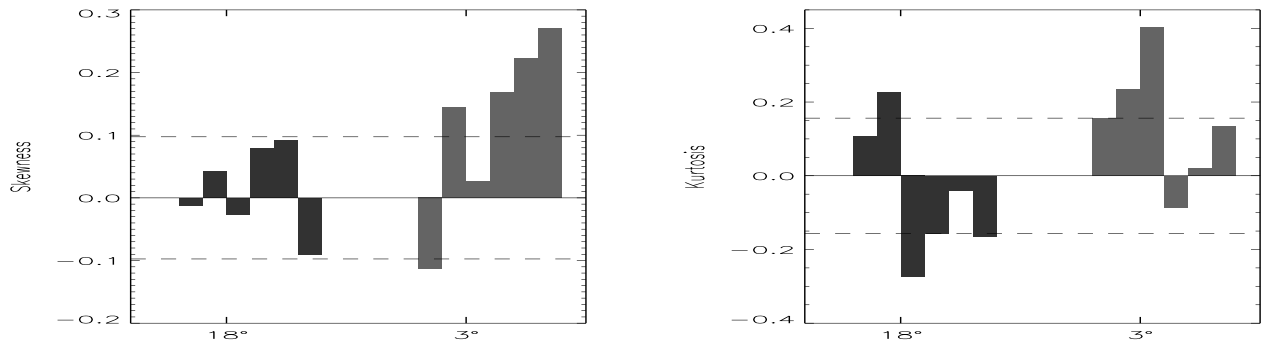}}
\caption{Skewness (left) and kurtosis (right) for all six
$18^{\circ}$ and $3^{\circ}$ maps.  The lines show the $1\sigma$ value for the
ensemble of Gaussian maps.}\label{moments}
\end{figure*}

We also computed the skewness
\begin{equation}
\gamma_1 = \frac{\frac{1}{N_{pix}}\sum_i{(T_i-\overline{T})^3}}{\sigma^3}
\end{equation}
and the kurtosis
\begin{equation}
\gamma_2 = \frac{\frac{1}{N_{pix}}\sum_i{(T_i-\overline{T})^4}}{\sigma^4} -3
\end{equation}
of the string maps and compared it to that of all our Gaussian maps. 
Figure~\ref{moments} shows the skewness and kurtosis for all the
maps as well as the $1\sigma$ contour value for the ensemble average of
Gaussian realizations.   This indicates a marginal positive $1\sigma$
skewness and kurtosis for the $3^\circ$ maps.   The absence of a strong
skewness in the $18^\circ$ maps seems to be at variance with the significant 
negative skewness $\gamma_1=-0.24$ found in the $7.2^\circ$ maps from 
late-time gravitational effects in Ref.~\cite{fraisse2008} (see also
the analytic estimates in~\cite{hindmarsh2009, RS09, yamauchi2010}).
Two possible explanations 
are apparent.  First, as mentioned above, the present 3D Einstein-Boltzmann simulations have an 
effectively lower resolution than the 2D flat-sky approximation maps,
so they cannot probe as far into the wings of the distribution.  We are missing 
the integrated effect of bispectrum triangles combining disparate large and 
small scales.   Second, the analytic modeling in Ref.~\cite{RS09} incorporates 
a causality or correlation-length
cut-off which prevents bispectrum and trispectrum contributions from
superhorizon scales, effectively flattening their accumulated amplitude below 
$l\approx 300-500$.   Without such a cut-off introduced through 
energy-momentum compensation, 
there may be a significant enhancement of the skewness.  

The tentative indications of positive skewness and kurtosis in the $3^\circ$ maps, 
however, are consistent with physical expectations.   On scales above $l>300$, we see
that there
is an important contribution from early scalars with strings generating wakelike
objects in the matter distribution with a positive skewness and kurtosis~\cite{avelino98b}.  
This will lead to CMB anisotropies with similar properties. It appears that this positive skewness 
will compete with and confuse the negative skewness predicted for the late-time gravitational effects
on intermediate scales.  In contrast, the kurtosis of these two contributions will 
sum positively, reinforcing the view that the trispectrum will prove a better 
discriminant of strings than the bispectrum~\cite{RS09}.   These preliminary 
results indicate that incorporating recombination effects will be important in any effort to 
constrain non-Gaussian strings signatures with WMAP or Planck data.   

\section{Conclusion}

We have computed maps of cosmic string induced CMB
fluctuations at various resolutions and we have extracted 
their angular power spectra.  We have demonstrated the 
importance of recombination effects for the power spectrum 
over a broad range of multipoles $200< l<2000$.    We have 
also shown that the resulting maps are remarkably Gaussian, 
though with potential deviations which are worthy of closer
investigation as testable string signatures in the CMB.   

\section*{Acknowledgments}

Numerical simulations were performed on COSMOS, the SGI Altix 4700,
owned by the United Kingdom Cosmology Consortium, funded by STFC, SGI and
Intel. We were also supported by STFC Grant No. ST/P002998/1. 
We are very grateful to Donough Regan, Carlos Martins, and Eiichiro Komatsu for
useful discussions about this work.  We thank Michele Liguori for his code
to generate Wigner 3j symbols. 

\bibliography{refs,myrefs,unpub}

\end{document}